\documentstyle[12pt]{article} 
\begin{document}

\renewcommand{\theequation}{\thesection .\arabic{equation}}
\newcommand{\beq}{\begin{equation}}
\newcommand{\eeq}{\end{equation}}
\newcommand{\beqn}{\begin{eqnarray}}
\newcommand{\eeqn}{\end{eqnarray}}
\newcommand{\slp}{\raise.15ex\hbox{$/$}\kern-.57em\hbox{$\partial
$}}
\newcommand{\lnA}{\raise.15ex\hbox{$/$}\kern-.57em\hbox{$A$}}
\newcommand{\lnB}{\raise.15ex\hbox{$/$}\kern-.57em\hbox{$B$}}
\newcommand{\bP}{\bar{\Psi}}

\begin{titlepage}

\rightline{UB-ECM-PF 93/9}

\begin{center} {\large{\bf Thirring Model with Non-conserved Chiral
Charge}} \end{center} \vspace{1.5cm} \begin{center} D.Cabra$^{a,b}$,
E.Moreno$^{c}$ and C.Na\'on$^{a,b}$ \end{center} \vspace{2cm}

ABSTRACT:{\small We study the Abelian Thirring Model when the fermionic
fields have non-conserved chiral charge: $\Delta {\cal Q}_5 =N$.
One of the main features we find for this model is the dependence of the
Virasoro central charge on both the Thirring coupling constant and $N$.
We show how to evaluate
correlation functions and
in particular we compute
the conformal dimensions for fermions and fermionic bilinears, which depend
on the fermionic chiral charge. Finally we build primary fields with arbitrary
conformal weight.}

\vspace{5cm}

\noindent --------------------------------

\noindent $^a$ {\footnotesize Depto. de F\'\i sica.  Universidad
Nacional de La Plata.  CC 67, 1900 La Plata, Argentina.}

\noindent $^b$ {\footnotesize Consejo Nacional de Investigaciones
Cient\'\i ficas y T\'ecnicas, Argentina.}

\noindent $^c$ {\footnotesize Departament d'Estructura i Constituents de
la Mat\`eria, Universitat de Barcelona, Diagonal 647, 08028, Barcelona,
Spain.}

\end{titlepage}

\section{Introduction}

\ \indent After the pioneering work of Belavin, Polyakov and Zamolodchikov
\cite{BPZ}, one of the most important steps towards the solution of 2D
conformal field theories was given by Dotsenko and Fateev \cite{DF} (DF).
These authors considered the Coulomb Gas Model with a non-trivial background
charge at infinity. In this way they got a conformal anomaly as well as
conformal weights that depend on this charge. With these basic ingredients
they developed a systematic procedure to evaluate a general $n$-point
correlation function of primary fields. Following their technique the operator
algebra of non-trivial critical 2D systems such as the $q$-component Potts
model \cite{Potts} and $O(N)$ model were obtained \cite{on}.

The success of this approach led us to explore the possibility of implementing
a similar idea but using a fermionic model as starting point. Thus, instead
of the bosonic Coulomb gas system, in this paper we consider a modified
version of the well-known Thirring model \cite{Thirring}. In order to introduce
the analogous of DF's charge at infinity, we spoil the conservation of the
chiral charge.

Our intention is to give an alternative formulation that could be useful
in the identification of the discrete variables in a given statistical model
with continuum fields.

A direct connection with statistical systems is lacking for an arbitrary
Conformal Field Theory, except for some trivial cases such as the Ising
Model, which is directly connected with a Majorana fermion via a Jordan Wigner
transformation \cite{Ising}. (The Ashkin-Teller \cite{AT} and Baxter \cite{Ba}
models have been also related to the usual Thirring model.)

This approach could facilitate the non-trivial task of relating lattice
and field theoretical models via a Jordan-Wigner-like transformation.

In this paper we study the Abelian Thirring model in the case in which the
chiral charge is not conserved. We found that the Virasoro central charge
(VCC) of the resulting model
depends on the ``topological charge" and gives rise to a family of conformally
invariant field theories with VCC given by:
\beq
c=1+3\frac{N^2\pi}{g^2}
\eeq
It is worthwhile noting the dependence of $c$ on the Thirring coupling
constant, which appears as a purely topological effect.

In the region in which the model is stable (which corresponds to $g^2>0$)
this charge is greater than one, and hence the series (1.1) does not include
for example Minimal Unitary models.

However one can study the region of negative couplings and in that case
our model could be useful in the study of non-unitary statistical models
like Scaling Lee-Yang model \cite{Fi}, etc.

Our model could be also relevant in the study of superconformal minimal
models, for those models in the series for which the VCC is in the interval
$[1,3/2]$ and other minimal unitary theories with additional symmetry, like
$Z_3$ symmetry, Parafermionic theories, etc.

On the other hand, our formulation allows to obtain the primary fields in
a simple way in contrast with the extremely complicated form that they take in
the bosonic and fermionic formulation of coset models \cite{GK}.

The paper is organized as follows: in Section II we present the model and
show how to evaluate the VCC. Section III is devoted to the study of
correlation
functions and the evaluation of the conformal dimensions of the primary
fields.

As a byproduct we obtain the conformal weights for some special bilinear
fermionic operators. This allows us to obtain the field theoretical description
of Baxter-like models.

\setcounter{equation}{0}

\section{The Model}

\ \indent Let us start with the euclidean Lagrangian:

\beq {\cal L}=
-\bar{\Psi}i\slp\Psi-\frac{1}{2}g^2(\bar{\Psi}\gamma_{\mu}\Psi)^2 ,
\label{1.1} \eeq which is invariant under the global abelian and abelian
chiral groups:  \beqn \Psi &\rightarrow& e^{i\alpha}\Psi , \nonumber \\
\Psi &\rightarrow& e^{\gamma_5\beta}\Psi , \label{1.2} \eeqn with
$\alpha~,\beta ~ \epsilon ~R$.

Our conventions for $\gamma$-matrices are:
\beqn
\{\gamma_{\mu},\gamma_{\nu}\}=2\delta_{\mu\nu}~~~~,~~~~\gamma_5
=i\gamma_0\gamma_1 ,\nonumber \\
\gamma_{\mu}\gamma_5=i\epsilon_{\mu\nu}\gamma_{\nu}
{}~~~~,~~~~\epsilon_{01}=1.
\label{1.3}
\eeqn

In the Path-Integral formulation of the theory we can eliminate
quartic
self interactions introducing an auxiliary field through the
identity:
\beq
exp\left(\frac{1}{2}g^2\int
d^2x(\bar{\Psi}\gamma_{\mu}\Psi)^2\right)=
\int DA_{\mu}exp\left(-\int d^2x (-g\bar{\Psi}\gamma_{\mu}\Psi
A_{\mu}+
\frac{1}{2}A_{\mu}A_{\mu})\right).
\label{1.4}
\eeq

Then the partition function of the model is given by:
\beq
{\cal Z}=\int D\bar{\Psi}D\Psi DA_{\mu} exp\left(-\int
d^2x(\bP(i\slp)\Psi+
g\bP\gamma_{\mu}\Psi A_{\mu}-\frac{1}{2}A_{\mu}A_{\mu})\right) .
\label{1.5}
\eeq

In this form we can easily study the theory in a non-trivial chiral
sector.  In fact, the field $A_{\mu}$ is coupled minimally with the
fermion current and has a natural interpretation as a gauge connection.
Therefore a gauge configuration carrying a winding number $N$ ,

\beq
\oint A_{\mu} dx^{\mu} =2\pi N.
\label{1.6}
\eeq
will produce a break in the chiral symmetry. It is easy to show
that the violation of the chiral charge is equal to $N$,

\beq
\Delta {\cal Q}_5 =N
\label{1.8}
\eeq
imposing the selection rule: {\it the transition amplitudes
between states
are non-zero only if the variation of the chiral charge is equal
to $N$.}

In order to study the partition function (\ref{1.5}) it is useful to
decompose the auxiliary field as \cite{BC}:

\beq
A_{\mu}=A_{\mu}^{(Cl)}+a_{\mu} ,
\label{1.11}
\eeq
where $A_{\mu}^{(Cl)}$ is a fixed background configuration which
carries
the topology and $a_{\mu}$ takes into account quantum
fluctuations, i.e.
\beq
DA_{\mu} =Da_{\mu} .
\label{1.12}
\eeq

With this choice the Lagrangian in eq.(\ref{1.5}) takes the form:
\beq
{\cal L}= \bP i\slp \Psi +g\bP \gamma_{\mu} \Psi A_{\mu}^{(Cl)}+
g\bP \gamma_{\mu} \Psi a_{\mu}-\frac{1}{2}
A_{\mu}^{(Cl)}A_{\mu}^{(Cl)}
- a_{\mu}A_{\mu}^{(Cl)} -\frac{1}{2} a_{\mu}a_{\mu}.
\label{1.13}
\eeq
Of course, by construction this partition function is independent
of the
background configuration $A_{\mu}^{(Cl)}$ because any two
configurations
differ only in a shift in the quantum fluctuations.

We can decouple the quantum part of the auxiliary field
($a_{\mu}$) from
fermions making the following change of variables \cite{RS},
\cite{GSSS}:
\beqn
\Psi &=& e^{i\eta+\gamma_5\phi}\chi \nonumber \\
\bP &=& \chi e^{-i\eta+\gamma_5\phi} \nonumber \\
a_{\mu}&=&-\frac{1}{g}(\epsilon_{\mu\nu}\partial_{\nu}\phi-\partial_{\mu}\eta),
\label{1.14}
\eeqn
and choose the background configuration $A_{\mu}^{(Cl)}$
satisfying:
\beq
\partial_{\mu}A_{\mu}^{(Cl)}=0 .
\label{1.15}
\eeq

Then the Lagrangian can be written as:

\beq
{\cal L}=\bar{\chi}(i\slp+g\lnA^{(Cl)})\chi+\frac{1}{g}\phi
F^{(Cl)}
-\frac{1}{2g^2}\left(
(\partial_{\mu}\eta)^2+(\partial_{\mu}\phi)^2\right)
-\frac{1}{2}(A_{\mu}^{(Cl)})^2  ,
\label{1.16}
\eeq
where $F^{(Cl)}=\epsilon_{\mu\nu}\partial_{\mu}A_{\nu}^{(Cl)}$.

The jacobians associated with the change of variables
(\ref{1.14})
are given by:
\beqn
J_f&=&exp\left\{-\frac{1}{2\pi}\int
d^2x((\partial_{\mu}\phi)^2-2g\phi F^{(Cl)})\right\} \nonumber \\
J_b&=&det\left\{-\frac{1}{g^2}(\partial_0^2+\partial_1^2)\right\} ,
\label{1.17}
\eeqn
and then the partition function can be written as:
\beq
{\cal Z}= det(-\frac{1}{g^2}\Delta)\times
{\cal Z}_{fermions}[A^{(Cl)}] \times {\cal Z}_{\eta-boson} \times
{\cal Z}_{\phi -boson}[A^{(Cl)}],
\label{1.18}
\eeq
where:
\beq
{\cal Z}_{fermions}[A^{(Cl)}]
=\int D\bar{\chi}D\chi
exp\left(-\int d^2x \bar{\chi}(i\slp+g\lnA^{(Cl)})\chi\right),
\label{1.19}
\eeq

\beq
{\cal Z}_{\eta-boson}=\int D\eta
exp\left(-\int d^2x
\frac{1}{2g^2}(\partial_{\mu}\eta)^2\right) ,
\label{1.20}
\eeq
and
\beq
{\cal Z}_{\phi -boson}[A^{(Cl)}]=\int D\phi
exp\left(-\int d^2x
\left(\frac{1}{2\pi}+\frac{1}{2g^2}\right)(\partial_{\mu}\phi)^2
-\left(\frac{1}{g^2}+\frac{1}{\pi}\right)\phi F^{(Cl)}
\right) .
\label{1.21}
\eeq

{}From the expression (\ref{1.18}) for the partition function we
can see that
the fermionic and bosonic sectors are coupled only through the
background
configuration $A_{\mu}^{(Cl)}$.

Now let us compute the conformal anomaly of this theory.  Because of the
independence of the partition function (\ref{1.21}) on the particular
election of the monopole background, we can choose it in such a way all
the magnetic field be concentrated at infinity (the north pole of $S^2$,
after compactification).  With this election the three sectors,
fermions, bosons and ghosts in the partition function (\ref{1.18})
became separately conformal invariant on the complex plane ${\cal
C}=S^2-\{north\ pole\}$.

Consequently the Virasoro algebra of the whole theory is the sum of the
three Virasoro algebras of the composing sub theories.  We will see that,
for the bosonic and the fermionic subsectors, the effect of the coupling
with the background field is similar to the charge at infinity in the
Dotsenko and Fateev Coulomb gas approach to conformal field theory
Ref\cite{DF}.

Let us start with the fermionic theory (\ref{1.19}). Using the
anomaly
equation
\beq
\partial_{\mu}j_{\mu}^5 = \frac{1}{\pi}F^{(Cl)}
\label{aa}
\eeq
and the gauge condition (\ref{1.15}) we can construct a pair of
{\it
conserved} currents \cite{moreno}
\beq
{\bf J}_z=\chi^{\dagger}_L \chi_L - \frac{1}{\pi}A^{(Cl)}_z
\label{aa1}
\eeq
\beq
{\bf J}_{\bar z}=\chi^{\dagger}_R \chi_R +
\frac{1}{\pi}A^{(Cl)}_{\bar z}.
\label{ab}
\eeq

The Noether charges associated with these currents takes the form, in
radial quantization
\beq
{\cal Q}_z=\oint dz {\bf J}_z \label{ac0}
\eeq
\beq
{\cal Q}_{\bar z}=\oint d{\bar z} {\bf J}_{\bar z}.  \label{ac}
\eeq
But using
the monopole quantization condition (and the fact that the monopole has
support at infinity) we see that the effect of the background field is
to create a charge at infinity of value $N$.  This charge changes the
neutral charge condition of the holomorphic and anti-holomorphic vacuum
expectation values and consequently there is a shift in the conformal
dimension of the primary fields.  At the level of the energy-momentum
tensor, this shift is produced by the appearance of an extra term, which
takes into account the changes in the conformal dimensions.  The total
$T_{zz}$ component of the energy-momentum tensor takes the form

\beq
T=(\chi^{\dagger}_L\partial_z \chi_L-\partial_z \chi^{\dagger}_L\chi_L)+
N \partial_z(\chi^{\dagger}_L\chi_L)
\label{ad}
\eeq
and there is a similar equation for the $T_{{\bar z}{\bar z}}$ component.
Using eq.(\ref{ad}) we can easily compute the Virasoro central
charge of
this fermionic theory and we find
\beq
c_{fermions}=1-3N^2.
\label{ae}
\eeq

Now we can evaluate the contribution from the bosonic fields
$\eta$ and
$\phi$.
For the $\eta$-field one get the well-known result:
\beq
c_{\eta-boson}=1.
\label{1.25}
\eeq
(This can be seen making a harmless rescaling of the
$\eta$-field:
$\tilde\eta=\frac{\sqrt{\pi}}{g}\eta$).

In order to evaluate the contribution coming from the $\phi$
field
it is useful to make the following rescaling
\beq
\tilde \phi=\sqrt{1+\frac{\pi}{g^2}}\phi ,
\label{1.26}
\eeq
which leads the action to the form:
\beq
S[\phi]=\frac{1}{2\pi}\left(\int d^2x
\left(\partial_{\mu}\tilde\phi\right)^2 -
Q\int d^2x \tilde\phi F^{(Cl)}\right) ,
\label{1.27}
\eeq
where:
\beq
Q=\sqrt{1+\frac{\pi}{g^2}}.
\label{1.28}
\eeq
The analysis of this theory is closely similar to the fermionic
one. In
particular it is easy to convince oneself that the coupling with the
monopole
produces a background charge at infinity of value $NQ$
(Ref.\cite{CM}). Then
the modification that suffers the energy-momentum tensor is given
by
\beq
\Delta T=Q\partial^2_z\tilde\phi
\label{ba}
\eeq
and the conformal anomaly takes the form
\beq
c_{\phi-boson}=1+3N^2\left(1+\frac{\pi}{g^2}\right) .
\label{1.29}
\eeq
Finally we have to take into account the contribution coming from
the
Laplacian in equation (\ref{1.18}).
The determinant of the Laplacian operator can be exponentiated
using
anticommuting ghosts in the form:
\beq
det(-\frac{1}{g^2}\Delta)=\int D\bar\eta D\eta exp\left(
-\frac{1}{g^2}\int
d^2 x \bar\eta \Delta \eta \right).
\label{1.22}
\eeq

The corresponding conformal charge is:
\beq
c_{ghosts}=-2 .
\label{1.23}
\eeq

At this point we can evaluate the total central charge of the
model just
by adding the four independent contributions
eqs.(\ref{ae}),(\ref{1.25}),
(\ref{1.29}), and (\ref{1.23}) which yields to:
\beq
c_{total}=1+3\frac{\pi N^2}{g^2}.
\label{1.30}
\eeq

We see that the central charge of this model depends both on the
coupling
constant $g^2$ and on the topological charge $N$, this last being
connected
with the non-conservation of the chiral charge (eq.(\ref{1.8})).

It is worthwhile noting the resemblance of this result with the
one obtained
by Dotsenko and Fateev \cite{DF}. In this sense our result
could be
viewed
as a fermionic version of DF's construction. Two important
differences
should be stressed, however. Firstly, we have a conformal charge
$c~>~1$
whereas DF got $c~<~1$. As a consequence, the FQS series
\cite{FQS} is not
described by our model, at least for values of the coupling
constant $g^2$
which lead to a unitary theory. Finally, it is remarkable the
dependence
of the central charge on the coupling constant $g^2$, which
arises as a
topological effect.

\newpage
\setcounter{equation}{0}

\section{Correlation Functions and Primary Fields}

Due to the presence of the monopole in the fermionic partition
function
eq.(2.19) only chirality violating operators have a non-zero
vacuum
expectation value. For a given fermionic operator ${\cal O}$ its
vacuum
expectation value is zero unless the selection rule
\beq
[{\cal Q}_5,{\cal O}]=N {\cal O}
\label{a.1}
\eeq
holds, where $N$ is the chiral charge. It is easy to verify that
the fermion
bilinears
${\bar \Psi}_R \Psi_R$ are eigenfunctions of the chiral charge
operator with
eigenvalue $1$, and the bilinears ${\bar \Psi}_L \Psi_L$ are
eigenfunctions
of the chiral charge operator with eigenvalue $-1$. Then, for
$N>0$, a fermionic operator with a
non-trivial vacuum expectation value contains a number of
bilinears
${\bar \Psi}_R \Psi_R$ which exceeds in $N$ the number of
bilinears
${\bar \Psi}_L \Psi_L$ (for $N<0$ we change $R \leftrightarrow
L$).
After the decoupling
change of variables eq.(\ref{1.14}) the fermionic operator $\cal O$
will be
accompanied by  bosonic vertex operator
\beq
V=e^{\int {\bf j}(x) \phi(x)}
\label{a.2}
\eeq
with
\beq
{\bf j}(x)=\sum_i \alpha_i \delta (x-x_i)\ ;\ \ \ \sum_i
\alpha_i=-2N.
\label{a.3}
\eeq

Now let analyze the bosonic partition function eq.(\ref{1.21}).
The vertex
operator eq.(\ref{a.2}) adds a new term to the action
\beq
\Delta S_b=- \int {\bf j}(x) \phi (x) d^2x.
\label{a.4}
\eeq
Then after the rescaling (2.26) the bosonic effective action
reads
\beq
S_b=\frac{1}{2\pi}\int \left\{(\partial\phi)^2-\phi\left[Q F^{(Cl)}
+
\frac{2\pi}{Q}{\bf j} +\frac{2\pi}{Q}\mu\right]\right\}
\label{a.5}
\eeq
where we included an external source
$\mu$. ($Q$ was defined in eq.(\ref{1.28})).

The path integral over the constant zero mode of the Laplacian
operator
imposes, as usual, the neutrality charge condition:
\beq
\frac{1}{4\pi}\int\left( Q F^{(Cl)} + \frac{2\pi}{Q}{\bf j} +
\frac{2\pi}{Q}\mu \right)d^2x = N(Q-\frac{1}{Q})+
\frac{1}{2Q}\int \mu d^2x =0.
\label{a.6}
\eeq
Then, in order to have a non-trivial result for the vacuum
expectation value
we need the presence of the additional source $\mu$ satisfying
\beq
\int \mu d^2x= (Q^2-1)N=\frac{\pi}{g^2}.
\label{a.7}
\eeq
This contribution can be obtained in a natural way from  the term
\beq
e^{-\frac{1}{g^2}\int A_{\mu} l^{\mu}}
\label{a.8}
\eeq
which comes from a source term for the currents
$ $ in the original Lagrangian
(eq.(\ref{1.1})).

Then a non-trivial vacuum expectation has the form

\beq
<{\cal H}[{\bar \Psi},\Psi] e^{\int j_{\mu} l^{\mu}}>=\int D{\bar \chi}D\chi
D\phi
exp\left(\int \bar{\chi}(i\slp+g\lnA^{(Cl)})\chi\right)\times
\eeq

\beq
exp\left(
\frac{1}{2\pi}\int \left\{(\partial\phi)^2-\phi\left[Q F^{Cl} +
\frac{2\pi}{Q}
{\bf j} -\frac{\pi}{g^2Q}\epsilon_{\mu
\nu}\partial_{\mu}l_{\nu}\right]\right\}
\right){\cal H}[{\bar \chi},\chi]
\label{a.9}
\eeq
where ${\cal H}[{\bar \chi},\chi]$ is a fermionic composite
operator
satisfying the chiral selection rule eq.(\ref{a.1}). The
constraint
eq.(\ref{a.7}) is translated into the following condition for the
source
$l_{\mu}$
\beq
\frac{1}{2\pi}\int \epsilon_{\mu \nu}\partial_{\mu}l_{\nu}d^2x
=4\pi N,
\label{a.10}
\eeq
{\it i.e.}, $l_{\mu}$ is a monopole of charge $N$.

Because of the freedom in the election of the classical
background
$A_{\mu}^{Cl}$
in the total partition function eq.(\ref{1.18}) we can choose $
A_{\mu}^{Cl}=l_{\mu}$. Finally integrating the remaining
(non-zero modes)
bosonic degrees of freedom we obtain the vacuum expectation
value of
${\cal O}={\cal H}[{\bar \Psi},\Psi] e^{\int j_{\mu} l^{\mu}}$

\beq
<{\cal O}>=\int D{\bar \chi}\chi
exp\left(\int \bar{\chi}(i\slp+g\lnA^{(Cl)})\chi\right)
{\cal H}[{\bar \chi},\chi] \times
\label{a.11}
\eeq
\beq
exp\left(-\frac{1}{8\pi Q^2}\int (F^{Cl}[l(x)]+2\pi{\bf j} (x)
)\Delta^{-1}(x,y)(F^{Cl}[l(y)]+2\pi{\bf j} (y) d^2x d^2y)\right),
\eeq
where $\Delta^{-1}(x,y)$ is the Green function of the  Laplacian.

The next task of this section is to construct the primary
fields
of the model and evaluate their conformal weights. As usual, given
a conformal field $\phi_h$ we can evaluate its conformal dimension
from the O.P.E
with the energy-momentum tensor. Using the usual normalization in
CFT this
O.P.E. reads:

\beq
T(z)\phi_h(w)=\frac{h}{(z-w)^2}\phi_h(w)+\frac{1}{z-w}\partial_w
\phi_h(w)+r.t.
{}.
\label{3.3}
\eeq

In the factored version of the model this can be achieved by
evaluating
the O.P.E's in each sector separately since our basic fields
$\Psi$ can
be expressed in terms of the fermion fields $\chi$, the bosonic
fields
$\phi$ (both coupled only to the background) and the free boson
$\eta$
(see eq.(\ref{1.14})). (As usually happens in abelian theories
the ghost sector
appears completely decoupled).

We shall first consider the fermionic sector (\ref{1.19}). The
corresponding EMT
was given in eq.(\ref{ad}) and for the O.P.E's with the fermionic
fields
$\chi_R$,
$\chi_L$, $\bar\chi_R$ and $\bar\chi_L$ we get:

\beq
T_{\eta}(z)\chi_R(\bar w)=\bar{T}_{\eta}(\bar z)\bar\chi_R(w)=
T_{\eta}(z)\bar\chi_L(\bar w)=\bar{T}_{\eta}(\bar z)\chi_L(w)=0 ,
\label{3.4}
\eeq

\beqn
\bar{T}_{\eta}(\bar z)\chi_R(\bar w)&=&
\frac{1-N}{2}\frac{1}{(\bar z-\bar w)^2}
\chi_R(\bar w)+\frac{1}{(\bar z-\bar w)}
\partial_{\bar w}\chi_R(\bar w) , \nonumber \\
T_{\eta}(z)\bar\chi_R(w)&=&\frac{1-N}{2}\frac{1}{(z-w)^2}
\bar\chi_R(w)+\frac{1}{(z-w)}\partial_{w}\bar\chi_R(w) ,
\nonumber \\
T_{\eta}(z){\chi}_L(w)&=&\frac{1+N}{2}\frac{1}{(z-w)^2}{\chi}_L(w
)+
\frac{1}{(z-w)}\partial_{w}{\chi}_L(w) , \nonumber \\
\bar{T}_{\eta}(\bar z)\bar\chi_L(\bar
w)&=&\frac{1+N}{2}\frac{1}{(\bar z-\bar
w)^2}\bar\chi_L(\bar w)+\frac{1}{(\bar z-\bar
w)}\partial_{\bar w}\bar\chi_L(\bar w) .
\label{3.5}
\eeqn

Concerning the ${\eta}$-boson, we are interested in the conformal
dimension
of the vertex operator:
\beq
\Phi^{(\eta)}_{\alpha}(z,\bar z)\equiv :e^{i\alpha\eta(z,\bar
z)}: .
\label{3.6}
\eeq
{}From the O.P.E. with the corresponding EMT we obtain:
\beq
h=\bar h=\frac{g^2}{8\pi}\alpha^2
\label{3.7}
\eeq

Finally we evaluate the O.P.E. of the ${\phi}$-vertex operator:

\beq
\Phi^{(\phi)}_{\alpha}(z,\bar z)\equiv :e^{i\alpha\phi(z,\bar
z)}:,
\label{3.8}
\eeq
with the EMT given in eq.(\ref{ba}).
The result is:
\beqn
T_{\phi}(z)\Phi^{(\phi)}_{\alpha}(w,\bar w)&=&\frac{\alpha}{8}
(2N-\frac{\alpha}{1+\frac{\pi}{g^2}})\frac{1}{(z-w)^2}
\Phi^{(\phi)}_{\alpha}(w,\bar w)
+\dots , \nonumber \\
\bar{T}_{\phi}(\bar z)\Phi^{(\phi)}_{\alpha}(w,\bar
w)&=&\frac{\alpha}{8}
(2N-\frac{\alpha}{1+\frac{\pi}{g^2}})\frac{1}{(\bar z-\bar w)^2}
\Phi^{(\phi)}_{\alpha}(w,\bar w)+\dots
\label{3.9}
\eeqn

Now we can write down the conformal weights associated to the
original fermions
${\Psi}_R$, ${\Psi}_L$, ${\bP}_R$ and ${\bP}_L$ just by adding
the previous
fermionic contributions (\ref{3.4}), (\ref{3.5}) and setting
${\alpha}$=1 in (\ref{3.7}) and
(\ref{3.9}). Thus we obtain:

\beqn
h_{\Psi_R}&=&\frac{N}{4}-\frac{g^2}{8(g^2+\pi)}+\frac{g^2}{8\pi}
, \nonumber
\\
\bar
h_{\Psi_R}&=&\frac{N}{4}-\frac{g^2}{8(g^2+\pi)}+\frac{g^2}{8\pi}+
\frac{1-N}{2} , \nonumber
\\
h_{\bar{\Psi}_R}&=&\frac{N}{4}-\frac{g^2}{8(g^2+\pi)}+\frac{g^2}{
8\pi}+
\frac{1-N}{2} , \nonumber
\\
\bar
h_{\bar\Psi_R}&=&\frac{N}{4}-\frac{g^2}{8(g^2+\pi)}+\frac{g^2}{8\pi}
\label{3.10}
\eeqn

\beqn
h_{\Psi_L}&=&-\frac{N}{4}-\frac{g^2}{8(g^2+\pi)}
+\frac{g^2}{8\pi}+\frac{1+N}{2} , \nonumber
\\
\bar
h_{\Psi_L}&=&-\frac{N}{4}-\frac{g^2}{8(g^2+\pi)}
+\frac{g^2}{8\pi} , \nonumber
\\
h_{\bar{\Psi}_L}&=&-\frac{N}{4}-\frac{g^2}{8(g^2+\pi)}
+\frac{g^2}{8\pi} , \nonumber
\\
\bar
h_{\bar\Psi_L}&=&-\frac{N}{4}-\frac{g^2}{8(g^2+\pi)}
+\frac{g^2}{8\pi}+\frac{1+N}{2}
\label{3.11}
\eeqn

It is interesting to observe that the scaling dimensions given by
\ $\Delta = h+\bar h$
do not depend on the chiral charge $N$. On the contrary, the
spins
$S = h-\bar h$ are changed by the background. We have, for
instance,

\beq
\Delta_{\Psi_R}=\frac{1}{2}+\frac{g^2}{4\pi}-\frac{g^2}{4\pi(1+g^
2/\pi)} ,
\label{3.12}
\eeq
\beq
S_{\Psi_R}=\frac{1-N}{2}  .
\label{3.13}
\eeq

We also note that (\ref{3.12}) exactly coincides with the result
obtained in
Ref.\cite{FGS} where the fermionic correlator was computed in the
${N=0}$
sector.

We will now focus our attention on primary fields that are built
as fermionic
bilinears. These are the energy-density and crossover operators
which are
defined, respectively, as:
\beq
\varepsilon(z, \bar z)=\bar\Psi_R\Psi_R(z, \bar
z)+\bar\Psi_L\Psi_L(z, \bar z)
, \label{3.14}
\eeq
\beq
C_r(z, \bar z)=\bar\Psi_R\bar\Psi_L(z, \bar z),~~~{\bar C}_r(z, \bar z)
=\Psi_R\Psi_L(z,\bar z).
\label{3.15}
\eeq
(Strictly speaking the crossover operator is defined, for $N=0$, as the
sum ${\cal C}_r=C_r+{\bar C}_r$. However in the case at hand $N\neq 0$,
eventhough $C_r$ and ${\bar C}_r$ have the same scaling dimensions, their
spins are different and consequently they do not add to a primary field).
The conformal behavior of this operators is of particular
importance in
view of the close connection between the Thirring model and some
lattice
spin systems like the Ashkin-Teller Ref.\cite{AT} and Baxter
models
Ref.\cite{Ba}. These two systems can be roughly described as the
superposition
of two Ising lattices which interact through 4-spin couplings. In
the context
of the Ashkin-Teller model ${\varepsilon}(x)$ accounts for
temperature
fluctuations, whereas ${C}_r(x)$ describes the difference between
the
nearest-neighbor couplings of the two Ising variables
Ref.\cite{KDK}.

Combining adequately the conformal dimensions of the constituents
(fermions
and vertex operators) we have:
\beqn
h_{\varepsilon}&=&\bar{h}_{\varepsilon}=
\frac{1}{2}\left(\frac{1}{1+\frac{g^2}{\pi}}\right) , \nonumber
\\
h_{C_r}&=&=\bar{h}_{{\bar C}_r}=-\frac{N}{2}+\frac{1+\frac{g^2}{\pi}}{2} ,
\nonumber \\
\bar{h}_{C_r}&=&=h_{{\bar C}_r}=\frac{N}{2}+\frac{1+\frac{g^2}{\pi}}{2} .
\label{3.16}
\eeqn

It must be stressed that the scaling dimensions for the energy and the
crossover operators
are related as:
\beq
\Delta_{C_r}=\Delta_{{\bar C}_r}=1+\frac{g^2}{\pi}=\frac{1}
{\Delta_{\varepsilon}} .
\label{3.17}
\eeq

This relation had been conjectured in Ref.\cite{E}, \cite{K}
and was first derived
in Ref.\cite{DK} for the ${N=0}$ case (see also Ref.\cite{N} for
a proof
of this identity in the path-integral framework). Eq.(\ref{3.17})
shows that this
relation holds irrespective of the boundary conditions for the
fermionic
fields.

Since the theory at hand has ${c>1}$ we must be able to construct
primary
fields with arbitrary conformal weights. Apart from special cubic
and quartic
combinations of fermions, the unique way to obtain this result is
to deal
with non-local operators Ref.\cite{FMS} like:
\beq
\Phi_{\alpha}(z)\equiv :e^{i\alpha \int_{C_z} dw j(w)}: ,
\label{3.18}
\eeq
where:
\beq
j(z)=2\pi:\bP_R\Psi_L:(z)=2\pi:\bar\chi_R\chi_L:(z) ,
\label{3.19}
\eeq
with:
\beq
j(z)j(w)=-\frac{1}{(z-w)^2} ,
\label{3.20}
\eeq
and $C_z$ is a complex path starting at zero and ending at $z$.

In order to calculate the O.P.E. of $\Phi_{\alpha}(w)$ with the
EMT it is useful to express the last one in the Sugawara form:

\beq
T(z)=-\frac{1}{2}:j(z)j(z):+\frac{iN}{2}:\partial_zj(z): .
\label{3.21}
\eeq

Using this expression for the EMT together with (\ref{3.20}) we
obtain:
\beq
h=\frac{\alpha(\alpha-N)}{2} .
\label{3.22}
\eeq

\section{Conclusions}

It is clear from the results of this paper that the conformal properties
of the Thirring model are modified when we consider a non-zero chiral
charge sector.  The conformal anomaly, equal to 1 in the zero charge
sector, has an additional contribution proportional to the chiral charge
an inversely proportional to the coupling constant
$$
c=1+3\frac{\pi N^2}{g^2}
$$
This result is very similar to the conformal anomaly of the
Dotsenko and Fateev's modified Coulomb gas but with an opposite sign.
Here the ratio $\frac{N}{g^2}$ plays the role of the (imaginary) charge
at infinity in DF's work.  Of course such a modification implies a
change in the whole conformal algebra, and in particular a new set of
primary and descendants conformal fields.  For the particular case of
the fermions we found that only the spins are $N$-dependent, the
scaling dimensions are independent of the chiral charge.

As it was shown by Luther and Peschel \cite{8'} one can generalize the
Jordan-Wigner transformation for the spin operators on a lattice model to
the continuum case. This fact enables one to study the lattice Baxter model
in terms of the continuum Thirring model (for the $N=0$ case).
In this context the fermionic bilinears $\varepsilon$ (\ref{3.14}) and $C_r$
(\ref{3.15}) play a central role.

We have computed the conformal weights
for these operators, when the fermion fields associated to the corresponding
Jordan-Wigner transformations satisfy non-trivial boundary conditions.We
found that $h_{\epsilon}$ and $\bar h_{\epsilon}$ are not affected by having
N different from zero. On the other hand, the crossover operator does change
in this new situation: it splits down into two primary operators with equal
scaling dimensions but opposite spins. These results can be understood by
recalling that the energy density is a rotationally invariant operator while
the crossover is not.

The non-trivial conditions satisfied by the fermionic fields lead to
non-trivial constraints
between the spin variables in the related lattice model. (This can be seen
by using the Jordan-Wigner transformation
which relates the usual Thirring model with the Baxter model)
It would be interesting to try to generate
these constraints in the lattice by the addition of an interaction
between the spins.

As mentioned in the Introduction, the central charge that we have obtained
for a given "topological charge" $N$ can be matched with the VCC of certain
Superconformal Minimal Models (those for which the VCC is in the interval
$[3/2, 1]$) by a suitable choice of the coupling $g^2$. For those cases
it would be interesting to construct the superpartner of the EMT and study
the connection with the realization of Superconformal invariance in related
lattice models.

The addition of a mass term to the Lagrangian (2.1) corresponds to a
perturbation with an operator of dimension $h=\bar h=1/2 (1/1+g^2/\pi)$.
In that case one can study the Renormalization Group flow between two adjacent
conformal field theories by evaluating the Zamolodchikov's $c$-function
\cite{Zam}.

\section{Acknowledgements}
E.F.M. thanks J. Soto and. J.I. Latorre for usefull comments.
D.C.Cabra and C. Na\'on are partially supported by CONICET.
E.F.M. acknowledges a Post-Doctoral fellowship from the Ministerio de
Educaci\'on y Ciencia. Partially supported by CICYT project AEN90-0033.

\end{document}